\documentclass[aps,prl,twocolumn]{revtex4}
\usepackage{amsmath}
\usepackage{amssymb}
\usepackage{graphics}

\newcommand\Schr{Schr\"odinger\,}

\newcommand\pr{^\prime}

\newcommand\e{\mathrm{e}}

\newcommand\px{\frac{\partial}{\partial x}}
\newcommand\pp{\frac{\partial}{\partial p}}
\newcommand\pptwo{\frac{\partial^2}{\partial p^2}}

\newcommand\ro{\hat\rho}
\newcommand\Ho{\hat H}
\newcommand\xo{\hat x}
\newcommand\po{\hat p}
\newcommand\Om{\Omega}
\newcommand\om{\omega}
\newcommand\Dth{D_\mathrm{th}}
\newcommand\Dsp{D_\mathrm{sp}}
\newcommand\Dm{D_\mathrm{m}}
\newcommand\DP{\mathrm{DP}}
\newcommand\CSL{\mathrm{CSL}}
\newcommand\DDP{D_\DP}
\newcommand\DCSL{D_\CSL}
\newcommand\DTsp{\Delta T_\mathrm{sp}}
\newcommand\DTm{\Delta T_\mathrm{m}}
\newcommand\dTm{\delta T_\mathrm{m}}
\newcommand\DTSQL{\Delta T_\mathrm{SQL}}
\newcommand\DTDP{\Delta T_\DP}
\newcommand\DTCSL{\Delta T_\CSL}
\newcommand\sDP{\sigma_\DP}
\newcommand\sCSL{\sigma_\CSL}
\newcommand\lCSL{\lambda_\CSL}

\begin{document}
\title{Testing spontaneous wave-function collapse models on classical mechanical oscillators}  
\author{Lajos Di\'osi}
\email{diosi.lajos@wigner.mta.hu}
\homepage{www.rmki.kfki.hu/~diosi}
\affiliation{Wigner Research Center for Physics, 
H-1525 Budapest 114. P.O.Box 49, Hungary}
\date{\today}

\begin{abstract}
We show that the heating effect of spontaneous wave-function collapse models
implies  an experimentally significant increment $\DTsp$ of equilibrium temperature in a mechanical oscillator. 
The obtained form $\DTsp$ is linear in the oscillator's relaxation time $\tau$ and independent of the mass. 
The oscillator can be in a classical thermal state, the effect $\DTsp$ is classical  for a wide range of frequencies and
quality factors. We note that the test of $\DTsp$ does not necessitate quantum state monitoring but tomography. 
In both gravity-related (DP) and continuous spontaneous localization (CSL) models the strong-effect edge of their
parameter range can be challenged in existing experiments on classical oscillators. 
For the CSL theory, the conjectured highest collapse rate parameter values become immediately constrained
by evidences from current experiments on extreme slow-ring-down oscillators.
\end{abstract}
\maketitle
Spontaneous collapse models \cite{Basetal13} suggest that large spatial superpositions of quantum states 
of massive degrees of freedom, also called \Schr  Cat states,  
decay at (model dependent) universal rates. These models, the particular
gravity-related (or DP) model \cite{Dio8789,Dio13,Dio14ab,Pen96} and the continuous spontaneous
localization (CSL) model \cite{GhiPeaRim90,Adl07} predict the progressive violation 
of the quantum mechanical superposition principle for massive degrees of freedom. 
For atomic degrees of freedom this violation
is irrelevant while for massive degrees of freedom it becomes significant though 
usually masked by the environmental noise. The preparation of \Schr Cat states is extremely
demanding hence the direct experimental test of spontaneous collapse
has not yet been achieved despite relentless efforts, see, e.g., 
\cite{Maretal03,Vanetal03,Rom11,LiKheRai11,Pepetal12,YinGerLi13,JuffUlbLi13}, 
and \cite{ArnHor14,AspKipMar13} for the state-of-the-art. 
Quite recently, Bahrami et al. \cite{Bahetal14} suggested a different approach, not requesting
laboratory \Schr Cat states. Nimmrichter et al. \cite{NimHorHam14} discuss the optomechanical sensing
of spontaneous momentum diffusion caused by collapse models.
We further elucidate and simplify these considerations and come to new results. 
We emphasize that momentum diffusion is classical and this facilitates the mathematical treatment, 
theoretical insight and experimental proposals.  Currently available mechanical oscillators of extreme long ring-down time, 
e.g.: in the Ref. \cite{Matetal13} by Matsumoto et al.,  are immediately capable of sensing spontaneous heating if
it exists with the strongest proposed rates.

Spontaneous collapse models \cite{Basetal13} impose spontaneous kinetic energy increase at constant rate proportional to the 
spontaneous collapse rate. This spontaneous heating is independent of the quantum state
which can be a classical state,  it need not to be a \Schr Cat state for being spontaneously heated.

While spontaneous collapse is a genuine quantum effect,
spontaneous heating is not. This we exploit in our work, an elementary (non-quantum) calculation yields 
the spontaneous increment $\DTsp$ of the equilibrium temperature $T$ of damped mechanical oscillators.
Full quantum calculations can be safely replaced by classical calculations as long as the oscillator
remains in the classical domain.  Most surprisingly, it turns out that in the classical domain
the current laboratory technique is already capable to test the spontaneous collapse models.  

{\it Spontaneous heating in oscillators.}  
Let us consider a quantized oscillator of mass $m$ and frequency $\Om$, with Hamiltonian
\begin{equation}\label{H}
\Ho=\frac{\po^2}{2m}+\frac{1}{2}m\Om^2\xo^2.
\end{equation}
If the mass is subject to spontaneous collapse, the density matrix $\ro$ satisfies the following master equation: 
\begin{equation}\label{ME}
\frac{d\ro}{dt}=\frac{-i}{\hbar}[\Ho,\ro]-\frac{\Dsp}{\hbar^2}[\xo,[\xo,\ro]]
\end{equation}
where $\Dsp$ governs the strength (rate) of spontaneous decoherence.
This  $\xo$-decoherence is observable:  it is simply equivalent with  $\po$-diffusion 
of diffusion constant $\Dsp$. 

From now on and through our work, we assume that the oscillator is in
the classical domain. Therefore we can describe it by the classical Liouville density $\rho(x,p)$ 
and the quantum master equation \eqref{ME} can be replaced by the Liouville equation
\begin{equation}\label{LE}
\frac{d\rho}{dt}=\{H,\rho\}+\Dsp\pptwo\rho.
\end{equation}
$H(x,p)$ is the classical Hamilton function of the oscillator, the Poisson
bracket  $\{H,\rho\}$ stands for $(p/m)\px\rho-m\Om^2q\pp\rho$ . In a realistic situation, 
the mechanical oscillator is in a thermal environment of temperature $T$, which will modify the Liouville equation: 
\begin{equation}\label{FPE}
\frac{d\rho}{dt}=\{H,\rho\}+\Dsp\pptwo\rho+\eta\pp p\rho+\Dth\pptwo\rho,
\end{equation}
where $\eta$ is the damping rate of oscillations and $\Dth=\eta mk_BT$ is the constant of thermal
momentum-diffusion. With $\Dsp=0$ we would get
the classical Fokker-Planck equation whose stationary solution is the
Gibbs canonical distribution $\mathcal{N}\exp(-H/k_BT)$. It is trivial to see
that with $\Dsp>0$ the stationary solution is the Gibbs canonical
distribution  
\begin{equation}\label{Gibbs}
\rho_\infty(x,p)=\mathcal{N}\exp{\left(-\frac{H(x,p)}{k_BT\pr}\right)}
\end{equation}
at the higher temperature
\begin{equation}\label{Teff}
T\pr=\left(1+\frac{\Dsp}{\Dth}\right)T\equiv T+\DTsp.
\end{equation}
This result can be interpreted as the extension of the Einstein-Smoluchowski relationship
$\Dth=\eta m k_B T$ for $\Dth+\Dsp=\eta mk_B T\pr$, supported by the underlying Fokker-Planck equation.
 
The increment $\DTsp>0$ over the environmental temperature $T$ is the contribution of spontaneous heating, 
this is the very  observable quantity that we wish to test. From Eq. \eqref{Teff}, we can express it as
\begin{equation}\label{DT}
\DTsp=\frac{\Dsp}{m k_B}\tau, 
\end{equation}
where $\tau=1/\eta$ will stand for the (energy) relaxation time  of the oscillator.
Our classical description is valid as long as the spontaneous heating concerns many
quanta of the oscillator:
\begin{equation}\label{HighDT}
k_B\DTsp\gg\hbar\Om.
\end{equation}

{\it Measurement.} Since we restrict ourselves for the classical domain \eqref{HighDT} of spontaneous
heating $\DTsp$, a single-shot classical (or quantum) measurement of precision $\dTm$ would detect $\DTsp$ 
provided $\dTm\lesssim\DTsp$. If this condition does not hold, we repeat the same measurement
many times, like in quantum state tomography.  Observe that quantum state monitoring is not necessary,
tomography is the more suitable means to detect spontaneous temperature increase of the previously prepared 
equilibrium oscillator state. Cumulative precision of tomography is not limited quantum theoretically.

For completeness, nonetheless, let us recapitulate the features of monitoring which is usually accompanied by   
some classical and/or quantum noise (back-action). We characterize this back-action by a further diffusion constant $\Dm$. 
The complete Liouville equation \eqref{FPE} reads:
\begin{equation}\label{FPEm}
\frac{d\rho}{dt}=\{H,\rho\}+\eta\pp p\rho+\left(\Dsp+\Dth+\Dm\right)\pptwo\rho.
\end{equation}
Suppose we start to measure the temperature of the oscillator at $t=0$.
The initial state of the oscillator is the Gibbs state \eqref{Gibbs} of temperature
$T+\DTsp$. When the `thermometer' is switched on, the measurement noise starts to
heat the oscillator towards the new stationary Gibbs state of temperature increased by
\begin{equation}\label{DTm}
\DTm=\frac{\Dm}{ m k_B}\tau.
\end{equation}
Trivial dynamics of heating follows from Eq.  \eqref{FPEm} in the limit $\eta\ll\Om$:
\begin{equation}\label{}
T\pr(t)=T+\DTsp+(1-\e^{-t/\tau})\DTm.
\end{equation}
Observe that the temperature effect of back-action is gradually reaching its steaty state value. 
Back-action can be ignored for times much shorter than $\DTm/\DTsp$ times $\tau$. 

There is no fundamental limitation on the measurement precision (fluctuations) $\dTm$ in the classical domain.
There is a quantum tradeoff between the spectral components of $\dTm$ and $\DTm$ at a chosen frequency $\om$:
\begin{equation}\label{dTmDTm}
\dTm\DTm\geq\frac{\hbar^2}{4k_B^2}\frac{\vert\Om^2-\om^2+i\eta\om/2\vert^2}{\eta^2},
\end{equation}
as it follows from Refs. \cite{Cavetal80}, cf. also Ref. \cite{NimHorHam14}. The minimum of $\dTm+\DTm$ is achieved when
\begin{equation}\label{SQL}
\dTm=\DTm=\frac{\hbar}{2k_B}\frac{\vert\Om^2-\om^2+i\eta\om/2\vert}{\eta}\equiv\DTSQL
\end{equation}
which is called the standard quantum limit. This limitation concerns the steady state spectral componenent of the
precision and back-action, respectively. For monitoring duration much shorter than $\tau$ (yet sufficient to gather significant data on 
$\DTsp$) the back-action won't  influence the system, we can choose finer precisions $\dTm$ than $\DTSQL$. 

\begin{table}[h]
\begin{tabular}{|r|c|c|c|c|c|}
\hline
                     &$10^2$           &$10^3$                           &$10^4$                          &$10^5$                           &$10^6$ \\
\hline
$10^5$Hz&[$10^{-8}$K]  &[$10^{-7}$K]                 &[$10^{-6}$K]                 &$10^{-5}$K                 &  $10^{-4}$K\\
\hline
$10^4$Hz&[$10^{-7}$K]  &$10^{-6}$K                 & $10^{-5}$K                   & $10^{-4}$K                   &$10^{-3}$K\\
\hline
$10^3$Hz& $10^{-6}$K   &$10^{-5}$K                     & ${10^{-4}}$K                &$10^{-3}$K                    &$\mathbf{10^{-2}}$K\\
\hline
$10^2$Hz& $10^{-5}$K   &$10^{-4}$K                     &$10^{-3}$K                    &$\mathbf{10^{-2}}$K &$\mathbf{10^{-1}}$K\\
\hline
$10$Hz     & $10^{-4}$K   &$10^{-3}$K                     &$\mathbf{10^{-2}}$K &$\mathbf{10^{-1}}$K &$\mathbf{1}$K \\
\hline
$1$Hz       &  $10^{-3}$K   &$\mathbf{10^{-2}}$K &$\mathbf{10^{-1}}$K &$\mathbf{1}$K             &$\mathbf{10}$K\\
\hline
\end{tabular}
\caption{Magnitudes of spontaneous heating effect $\DTDP$ of the DP-model on classical oscillators are  
shown at currently available or nearly available combinations of frequencies $\Om$ and quality factors $Q$. 
The spatial resolution $\sDP=10^{-12}$ cm assumes the strongest effect, lattice constant is set to $500$ pm.
Data around the upper-left corner (it brackets) are not in the classical domain $k_B\DTDP\gg\hbar\Om$.  
Data above the millikelvin range are enhanced (typed in boldface) because their detection may not request millikelvin cooling or 
cooling at all.}  
\label{table:DTvsOmQ}
\end{table}

{\it Spontaneous heating: DP-model.} 
In the gravity-related spontaneous collapse model (DP-model), the spontaneous diffusion is proportional
to the Newton constant $G$. For the simple example of oscillator mass considered in \cite{NimHorHam14}: 
\begin{equation}\label{DDP}
\DDP=\frac{\hbar}{2}m\om_G^2
=\frac{\hbar}{2}m\frac{4\pi G\varrho}{3}\left(\frac{a}{2\sqrt{\pi}\sDP}\right)^3
\end{equation}
where  $\varrho$ is the  mass density, and $a$ is the lattice constant, while $\om_G$ is the effective parameter used by \cite{Dio13,Dio14ab}.
The spatial resolution $\sDP$ is the free parameter
of the DP-model, conjectured to be in the following range \cite{Dio13}:  
\begin{equation}\label{sigmaDP}
10^{-12} cm\lesssim\sDP\lesssim10^{-5} cm.
\end{equation} 
The expression \eqref{DDP} is valid for $\sDP\ll a$. In this range, $\DDP$ is independent of the shape of
the mass while it depends on its microscopic structure.
Using \eqref{DDP} for $\Dsp$, we can write \eqref{DT} as
\begin{equation}\label{DTDP}
\DTDP=\frac{\hbar\om_G^2}{2k_B}\tau,
\end{equation}
where $\om_G^2$ is read out from \eqref{DDP}. 
It is remarkable that $\DTDP$ does not depend on the mass $m$.

Now we assume the strongest possible DP-decoherence, i.e., we take the finest 
conjectured spatial resolution $\sDP=10^{-12}$ cm, also favored by particular arguments \cite{Dio13,Dio14ab}.
If the lattice constant is set to $a=5\times10^{-8}$ cm, for concreteness, we obtain $\om_G\approx1.3$ kHz for the effective parameter. 
The spontaneous heating  effect \eqref{DTDP} can be written as
\begin{equation}\label{DTDPnum}
\DTDP\approx\tau[s]\times4.0\times10^{-5}K.
\end{equation}
This is a convenient expression of the effect $\DTDP$ to discuss possible
choices of the frequency $\Om$ and the quality factor $Q=\Om\tau$ of the oscillator. The mass $m$
has, as we noticed before, canceled from $\DTDP$.

{\it Experimental implications.} Applying Eq. \eqref{DTDPnum} to a broad range of 
frequencies $\Om$ and quality factors $Q$, we calculated the spontaneous 
heating $\DTDP$ in Table \ref{table:DTvsOmQ}. 

The lesson is transparent. If $\DTDP\gg\hbar\Om/k_B$, and this is the case except for a few highest $\Om$ and lowest $Q$ examples (in brackets),
the DP-effect would prevent us from ground state cooling. This should be a significant detectable effect.
But we do not need to try ground state cooling, the heating effect $\DTDP$ equally shows up far from the ground state.
Low frequency oscillators with high quality factors are the
favorable testbed.   If the ring-down time $\tau=Q/\Om$ of the oscillator is chosen between $10^2$ s and $10^6$ s, 
the spontaneous heating $\DTDP$ scales between $1$ mK and $10$ K, respectively. This is a striking result.
It is clear that classical (non-quantum) 
`thermometers' of precision $\dTm\sim1$ mK should exist. Technically, nonetheless, we might need to operate the measurement device
in the quantum domain especially when the oscillator itself cooled and/or controlled via high precision quantum devices.
Even in this case the oscillator is assumed to stay away from its ground state since the effect $\DTDP$ is robust classical. 

Following Ref. \cite{NimHorHam14}, and for a selection of experiments considered therein,
we calculated the effect $\DTDP$, see Table \ref{table:DTvsEXPR}.
The experiments  \cite{Muletal08}   and  \cite{Matetal13}, both performed at room temperature $T=300$ K, might be the promising ones. 
On the one hand, cooling is a reserve of higher sensitivity  of detecting $\DTDP$. On the other hand, the experiment  \cite{Matetal13} even at room 
temperature must be sensitive to the $6.4$ K spontaneous warming up.
 
As we mentioned before, monitoring may be neither convenient nor sufficient for detection.
Let us consider the constraint \eqref{dTmDTm} at
the detection band around $\om=2\pi\times500$, yielding $\dTm\DTm=\DTSQL^2=(37$ K$)^2$. Such a standard quantum limit
$37$ K gives insufficient precision on the steady state, i.e.: in monitoring of duration much longer than $\tau=1.6\times10^5$ s. 
If we choose $\dTm=1$ K the duration of monitoring must be limited to the order of hundred seconds before the back-action reaches the range of $1$ K. 
This is obviously not the way to go in general. In this particular experiment measurement precisions below $1$ K are not available by standard quantum monitoring.  
A single-pulse measurement must be considered instead, where state preparation is followed by a single one-shot measurement and the preparation-detection
cycle is repeated many times.  

\begin{widetext}
\begin{table*}
\begin{tabular}{lcccccc}
{\bf System}                                           &$m$                     &$\Om/2\pi$ (Hz)     &$Q$                          &$T$ (K)    &$\DTDP$ (K)\\     
\hline
gravitational wave detector   \cite{Muletal08}         
                                                                   &$40$ kg               &$1$                             &$25000$                  &$300$     &$0.16$\\
suspended disc  \cite{Matetal13}                                   
                                                                   &$5$ mg                &$0.5$                          &$5\times10^5$      &$300$     &$6.4$\\
SiN membrane   \cite{PurPetReg13}                                   
                                                                   &$34$ ng               &$1.6\times10^6$   &$1100$                    &$4.9$        &[$4.4\times10^{-9}$]\\
aluminium membrane  \cite{Teuetal11}                      
                                                                   &$48$ pg               &$1.1\times10^7$   &$3.3\times10^5$  &$0.015$   &[$1.9\times10^{-7}$]\\
\end{tabular}
\caption{Spontaneous heating $\DTDP$ for the selection of opto-mechanical setups quoted in \cite{NimHorHam14}.
Values $\DTDP$ are calculated from Eq. \eqref{DTDP}, assuming the largest  spontaneous decoherence rates considered for the time being, 
corresponding to $\om_G=1.3$kHz. Two of the data (in brackets) are not in the classical domain $k_B\DTDP\gg\hbar\Om$.}  
\label{table:DTvsEXPR}
\end{table*}
\end{widetext}

{\it CSL-model.} In the CSL model the diffusion constant is proportional to the rate parameter $\lCSL$.
For the perpendicular momentum diffusion of a disk of thickness $d$ it reads 
\begin{equation}\label{DCSL}
\DCSL=\lCSL\frac{\hbar^2}{m_0^2}4\pi\sCSL^2\frac{\varrho m}{d},
\end{equation}
where $m_0$ is the standard atomic unit.
The value of the CSL collapse rate parameter has been constrained by a lower \cite{GhiPeaRim90} and an upper estimate \cite{Adl07}, cf. also \cite{Basetal13}: 
\begin{equation}\label{limlambda}
2.2\times10^{-17} Hz\lesssim\lCSL\lesssim2.2\times10^{-8\pm2} Hz.
\end{equation}
Using $\DCSL$ \eqref{DCSL} for $\Dsp$ in \eqref{DT} yields 
\begin{equation}\label{DTCSL}
\DTCSL=\lCSL\frac{\hbar^2}{m_0^2 k_B}4\pi\sCSL^2\frac{\varrho}{d}\tau.
\end{equation}
Note that the shape (thickness)  of the oscillator matters, the mass $m$ does not.

Suppose the strongest CSL decoherence rate from the range \eqref{limlambda},  let's take 
the estimate $\lCSL=2.2\times10^{-8\pm2}$ Hz \cite{Adl07}.
Using this value in \eqref{DTCSL} we obtain
\begin{equation}\label{DTCSLnum}
\DTCSL\approx\tau[s]\frac{\varrho[g/cm^3]}{d[cm]}\times3.2\times 10^{-6\pm2}K.
\end{equation}
Recall that $d\gg\sCSL=10^{-5}$ cm, 
hence the strongest heating effect is achieved when $d\approx\sCSL$, leading to 
\begin{equation}\label{DTDPCSLnum}
\DTCSL\approx\tau[s]\times6.2\times 10^{-1\pm2}K,
\end{equation}
where we kept $\varrho=2$ g/cm$^3$ as before. Comparing this result with \eqref{DTDPnum} we conclude that, in classical oscillators,
the strongest conjectured CSL effect $\DTCSL$ would exceed the strongest conjectured DP effect $\DTDP$ by at least two orders of magnitude.

Let us consider the $\Om=3.14$ Hz oscillator \cite{Matetal13},  also discussed in Ref. \cite{NimHorHam14} in the context of the CSL model.
Recall that the strongest DP-effect turned out to be $\DTDP=6.4$ K, cf. Table \ref{table:DTvsOmQ}.
This oscillator has the high quality factor $Q=5\times10^5$, the ring-down time is extreme long: $\tau=1.6\times10^5$ s.
The resonator is a $5$ mg disk of thickness $d=0.2$ mm, Eq. \eqref{DTCSLnum} yields the 
spontaneous heating  $\DTCSL=5.1\times10^{1\pm2}$ K, corresponding to the rates $\lCSL=2.2\times10^{-8\pm2}$, respectively.
Obviously the values $\lCSL\gtrsim10^{-8}$ are not compatible with the experiment and the values
$\lCSL\sim(10^{-9}-10^{-10})$  remain to be challenged.  
 
{\it Summary.} 
The so far hypothetic spontaneous wave-function collapse on  massive degrees of freedom possesses a complementary classical effect:
classical momentum diffusion. This produces a certain spontaneous increase $\DTsp$ of the equilibrium temperature.  
This typical classical effect must be testable classically, without facing the standard quantum limitations of sensing. 
Therefore we must get spontaneous diffusion in the cross hairs instead of spontaneous collapse.  
We have derived the spontaneous heating $\DTsp$ for mechanical oscillators in classical thermal state, only using the 
classical Einstein-Smoluchowski relation, and found that $\DTsp$ is proportional to the relaxation (ring-down) time,
is independent of the mass. Experimental implications become transparent for both leading models DP and CSL of
spontaneous collapse.  We conclude that currently available extreme low-loss mechanical oscillators 
can already confirm the presence of spontaneous diffusion if its rate is close to the conjectured maximum.  
Alternatively, they enforce the update of the current constraints, cf. in Refs. \cite{Basetal13,FelTum12}, on collapse model's parameters. 
The requested measurement precisions $1$ mK - $1$ K may not be reached in standard steady state quantum monitoring. We suggested
that  state tomography will fit the demands.

\phantom{}
This work was supported by the Hungarian Scientific Research Fund under Grant No.  103917,  and by EU COST Actions MP1006, MP1209.

\end{document}